\documentclass{iaus}
\usepackage{graphicx}

\def\spose#1{\hbox to 0pt{#1\hss}}
\def\la{\mathrel{\spose{\lower 3pt\hbox{$\mathchar"218$}}
     \raise 2.0pt\hbox{$\mathchar"13C$}}}
\def\ga{\mathrel{\spose{\lower 3pt\hbox{$\mathchar"218$}}
     \raise 2.0pt\hbox{$\mathchar"13E$}}}
\def\msun{{\rm M}_{\odot}}

\def\citep#1{(\citealt{#1})}  

\def\day{\rm day}

\def\spitzer{{\it Spitzer}}

\title[Rotational evolution] 
{The rotational evolution of low-mass stars}

\author[Irwin \& Bouvier]   
{Jonathan Irwin$^1$ \and Jerome Bouvier$^2$}

\affiliation{$^1$Harvard-Smithsonian Center for Astrophysics, \\ 60 Garden Street MS-16, Cambridge, MA 02138, USA \\ email: {\tt jirwin -at- cfa.harvard.edu} \\
$^2$Laboratoire d'Astrophysique, Observatoire de Grenoble, \\ BP 53,
  F-38041 Grenoble C\'{e}dex 9, France}

\pubyear{2008}
\volume{258}  
\pagerange{}
\jname{The Ages of Stars}
\begin{document}

\maketitle

\begin{abstract}
We summarise recent progress in the understanding of the rotational
evolution of low-mass stars (here defined as solar mass down to the
hydrogen burning limit) both in terms of observations and modelling.
Wide-field imaging surveys on moderate-size telescopes can now
efficiently derive rotation periods for hundreds to thousands of
open cluster members, providing unprecedented sample sizes which are
ripe for exploration.  We summarise the available measurements,
and provide simple phenomenological and model-based interpretations of
the presently-available data, while highlighting regions of parameter
space where more observations are required, particularly at the lowest
masses and ages $\ga 500\ {\rm Myr}$.

\end{abstract}


\firstsection 

\section{Rotation period data}
\label{obs_section}

\begin{figure}[h]
\begin{center}
\includegraphics[width=3.2in,angle=270]{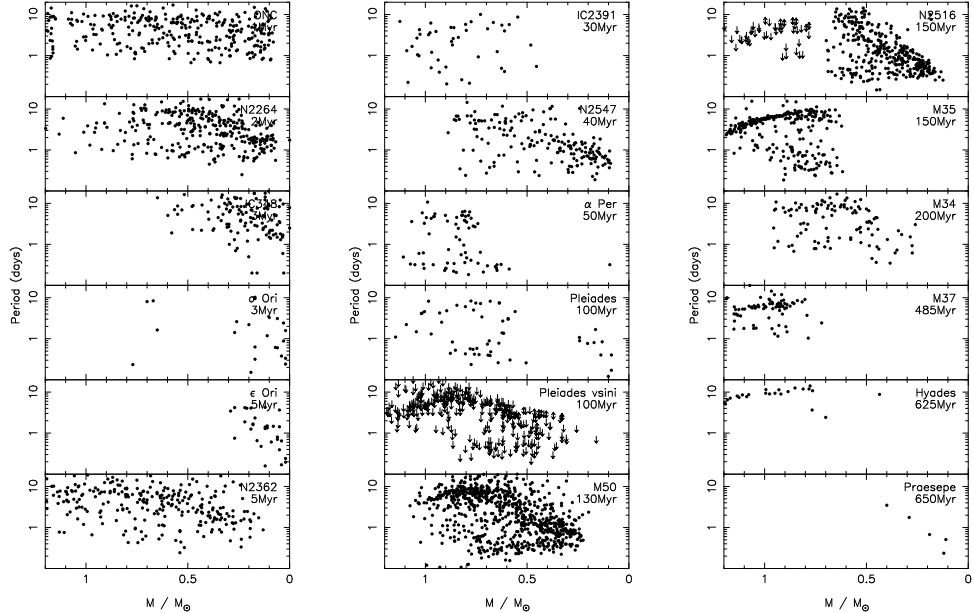}
\caption{Compilation of $3100$ rotation periods (and some $v \sin i$
  measurements) for stars with masses $M \la 1.2\ \msun$ in young
  ($\la 1\ {\rm Gyr}$) open clusters, from the literature.  Plotted in
  each panel is rotation period as a function of stellar mass for a
  single cluster.  The appropriate references for each panel are given
  in Table \ref{source_table}.  All the masses used in this
  contribution were computed using the I-band luminosities of the
  sources and the models of \cite{bcah98}, assuming values of the age,
  distance modulus and reddening for the clusters taken from the
  literature.}
\label{pmd_all}
\end{center}
\end{figure}

A compilation of most of the available rotation period (and some $v
\sin i$) measurements in open clusters for stars with masses $M \la
1.2\ \msun$ is shown in Figure \ref{pmd_all} (a list of references is
included in Table \ref{source_table}).  We plot rotation period
as a function of stellar mass, rather than using the more conventional
quantities of colour or spectral type on the horizontal axis, since
the diagram spans ages from the early pre--main-sequence (PMS), at
$\sim 1\ {\rm   Myr}$ in the ONC, to the main sequence (MS), and
neither colour nor spectral type are invariant for a given star over this age
range.  The masses are, of course, model-dependent, but the majority
of binning in mass and morphological examination used in this work
require only that the mass scale is {\it approximately} correct, which
should be reasonably well-satisfied by the PMS stellar evolution
tracks.  We use those of the Lyon group, from \cite{bcah98} and
I. Baraffe (private communication), throughout.

\begin{table}[hp]
\begin{center}
\begin{tabular}{lll}
\hline
\hline
Cluster  &Age  &Source(s) \\
         &(Myr) &         \\
\hline
\hline
ONC      &$1$ &\cite[Herbst et al. (2001,2002)]{herbst2001,herbst2002} \\
         &    &\cite{stassun99} \\
\hline
NGC 2264 &$2$ &\cite{lamm2005} \\
         &    &\cite{makidon2004} \\
\hline
IC 348   &$3$ &\cite{cohen2004} \\
         &    &\cite{little2005} \\
         &    &\cite{cieza2006} \\
\hline
$\sigma$ Ori &$5$ &\cite{se2004a} \\
\hline
$\epsilon$ Ori &$5$ &\cite{se2005} \\
\hline
NGC 2362 &$5$ &\cite{i2008b} \\
\hline
IC 2391  &$30$ &\cite{ps96} \\
IC 2602  &     &\cite{bsps99} \\
\hline
NGC 2547 &$40$ &\cite{i2008a} \\
\hline
$\alpha$ Per  &$50$ &\cite[Stauffer et al. (1985, 1989)]{aper1,aper2} \\
         &     &\cite{aper3} \\
         &     &\cite{aper4} \\
         &     &\cite[O'Dell et al. (1994, 1996)]{aper5,aper6} \\
         &     &\cite{aper7} \\
         &     &\cite{aper8} \\
         &     &\cite{aper9} \\
         &     &\cite{aper10} \\
         &     &\cite{aper11} \\
\hline
Pleiades &$100$ &\cite{ple1} \\
         &      &\cite{ple2} \\
         &      &\cite{ple3} \\
         &      &\cite[Prosser et al. (1993a,b,1995)]{ple4,ple5,ple6} \\
         &      &\cite{ple7} \\
         &      &\cite{t99} \\
         &      &\cite{se2004b} \\
\hline
Pleiades $v \sin i$  &$100$ &\cite{plev1} \\
         &      &\cite{plev2} \\
         &      &\cite{plev3} \\
         &      &\cite{plev4} \\
         &      &\cite{plev5} \\
         &      &\cite{plev6} \\
\hline
M50      &$130$ &\cite{i2009} \\
\hline
NGC 2516 &$150$ &\cite{i2007} \\
\multicolumn{1}{r}{$v \sin i$}         &      &\cite{ter2002} \\
\multicolumn{1}{r}{$v \sin i$}         &      &\cite{jeffries98} \\
\hline
M35      &$150$ &\cite{meibom2008} \\
\hline
M34      &$200$ &\cite{i2006} \\
\hline
M37      &$485$ &\cite{hartman2008} \\
\hline
Hyades   &$625$ &\cite{hya1} \\
         &      &\cite{ple6} \\
\hline
Praesepe &$650$ &\cite{se2007} \\
\hline
\hline
\end{tabular}

\caption{List of references for the panels in Figure \ref{pmd_all}.}
\label{source_table}
\end{center}
\end{table}

\begin{figure}[h]
\begin{center}
\includegraphics[width=3in,angle=270]{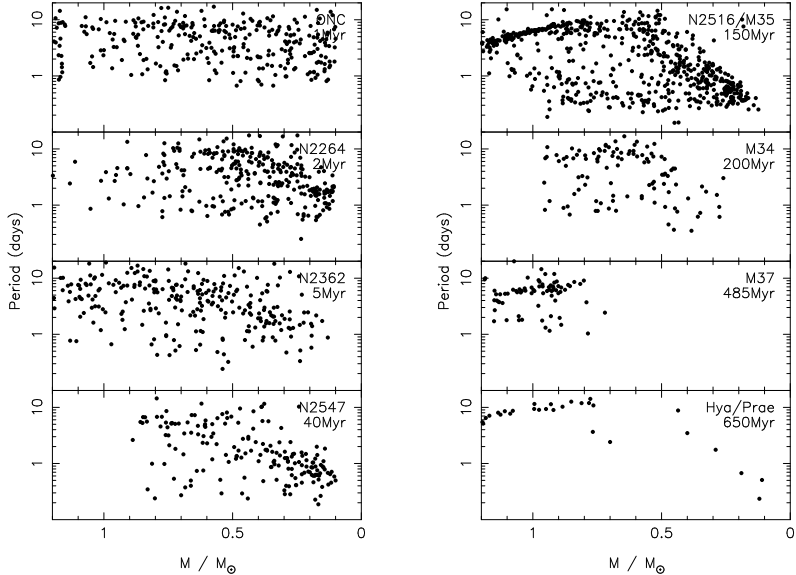}
\caption{Enlarged version of Figure \ref{pmd_all}, showing a few
  selected clusters which have large samples of rotation periods
  available.  NGC 2516 and M35 have been combined into a single panel,
  since they have essentially identical ages, as have the Hyades and
  Praesepe.}
\label{pmd_combined}
\end{center}
\end{figure}

An expanded version of Figure \ref{pmd_all}, omitting many of the more
sparsely sampled clusters and all the $v \sin i$ observations, is
shown in Figure \ref{pmd_combined}.  By examining the 
evolution of the morphologies of these diagrams, we can already draw
some (model-independent) conclusions regarding the evolution of the
rotation periods in these clusters.  For simplicity, we discuss two
broad mass ranges: stars close to solar mass (e.g. $0.9 \la M/\msun
\la 1.1$), and fully-convective stars ($M \la 0.4\ \msun$), which
represent the ``tail'' of the distribution which emerges especially in
the NGC 2547 and NGC 2516/M35 panels of Figure \ref{pmd_combined}.

Considering the solar mass stars first, it is clear that the behaviour
on the PMS (which lasts to $\sim 30\ {\rm Myr}$ for a solar mass star,
until it reaches the zero-age main sequence, hereafter ZAMS) and on
the MS up to the age of the Hyades, varies as a function of
rotation rate.  The slowest rotators in the ONC have periods of $\sim
10\ {\rm days}$.  The most basic prediction we can make for the
evolution of the rotation rate is to assume angular momentum is
conserved, in which case as the star contracts towards the ZAMS, it
should spin up.  However, for these slowly-rotating stars, the period
remains approximately constant all the way to NGC 2362, and has spun
up to $\sim 8\ {\rm days}$ by the age of NGC 2547 ($\sim 40\ {\rm
  Myr}$), at which point the contraction ceases.  Over the same age
range, the most rapid rotators gradually evolve from a period of $\sim
1\ {\rm day}$ at the ONC, to $\sim 0.6\ {\rm days}$ at NGC 2362 ($\sim
5\ {\rm Myr}$), and $\sim 0.2\ {\rm 
  days}$ at NGC 2547.  This follows closely the prediction from
stellar contraction.  We therefore conclude that there is some
mechanism removing angular momentum from only the slow rotators, to an
age of $\sim 5\ {\rm Myr}$, and a short time after this the angular
momentum losses cease, leaving the star free to spin up for a short
time until it reaches the ZAMS.  The net result of this is to yield a
wide range of rotation rates in the early-MS clusters such
as the Pleiades and M35.

By the age of the Hyades however, the rotation rates converge onto a
single well-defined sequence.  It is clear that
this process is well underway by the age of M37 ($\sim 485\ {\rm
  Myr}$).  There must therefore be another rotation-rate-dependent
angular momentum loss mechanism operating on the early-MS,
such that more rapid rotators lose more angular momentum, to drive all
the stars toward the same rotation rate at a given mass.

In contrast, the behaviour of the lowest mass stars is clearly rather
different.  On the PMS, these stars all appear to spin up rapidly, and
reach very rapid rotation rates at the ZAMS (for a $0.4\ \msun$ star,
the ZAMS is reached at $\sim 150\ {\rm Myr}$), and furthermore, the
maximum rotation period seen is a very strong function of mass (this
is most clear in the NGC 2516 diagram in Figure \ref{pmd_combined}).
This indicates that there is little rotation rate dependence in this
mass domain due to the similarities between the morphologies of the
diagrams especially from NGC 2264 to NGC 2516, and furthermore, that
these stars lose little angular momentum on the PMS, in contrast to
the slowly rotating solar mass stars.  Moreover, between NGC 2516 and
Praesepe, the limited quantity of data available suggest that there
is essentially no evolution of the rotation period, so little angular
momentum appears to be lost on the early-MS, again in
contrast to the solar-type stars.

We proceed in \S \ref{model_section} to examine the physical
interpretation of these observational results.

\section{Models of rotational evolution}
\label{model_section}

Recall from \S \ref{obs_section} that we needed to invoke two
mechanisms of angular momentum removal: one operating for $\sim 5-10\
{\rm Myr}$ on the PMS, and one operating on the main
sequence, with the property that it must produce a convergence in the
rotation rates between $\sim 100$ and $600\ {\rm Myr}$.

\subsection{Pre--main-sequence angular momentum losses}

For the first of these, the PMS angular momentum loss, the $\sim 5-10\
{\rm Myr}$ timescale suggests an obvious candidate: the circumstellar
discs that surround these stars in the earliest stages of their
evolution dissipate on these timescales (e.g. \cite[Haisch et
al. 2001]{haisch2001}).  The precise mechanism by which angular
momentum is removed due to the presence of a disc is still a matter of
debate, with the currently-favoured hypothesis being an
accretion-driven stellar wind (e.g. \cite[Matt \& Pudritz
2005]{mp2005}), or ``disc locking'' (e.g. \cite[K\"onigl
1991]{konigl91}; \cite[Collier Cameron et al. 1995]{cc95}).  For our
purposes, we shall simply treat the effect of the disc as holding the
angular velocity of the star constant (by removing angular momentum)
for the lifetime of the disc.

\begin{figure}[h]
\begin{center}
\includegraphics[width=1.6in,angle=270]{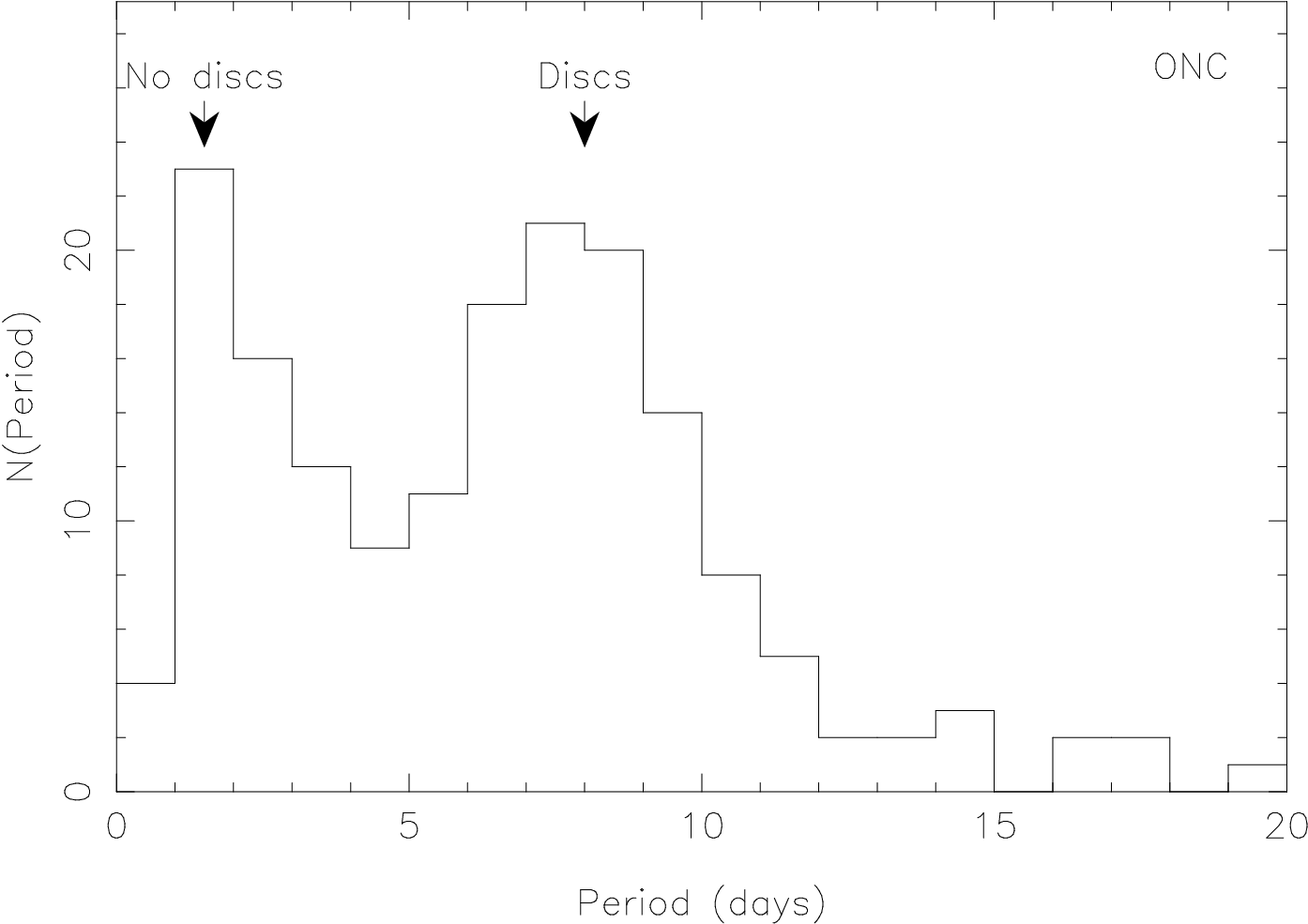}
\includegraphics[width=1.6in,angle=270]{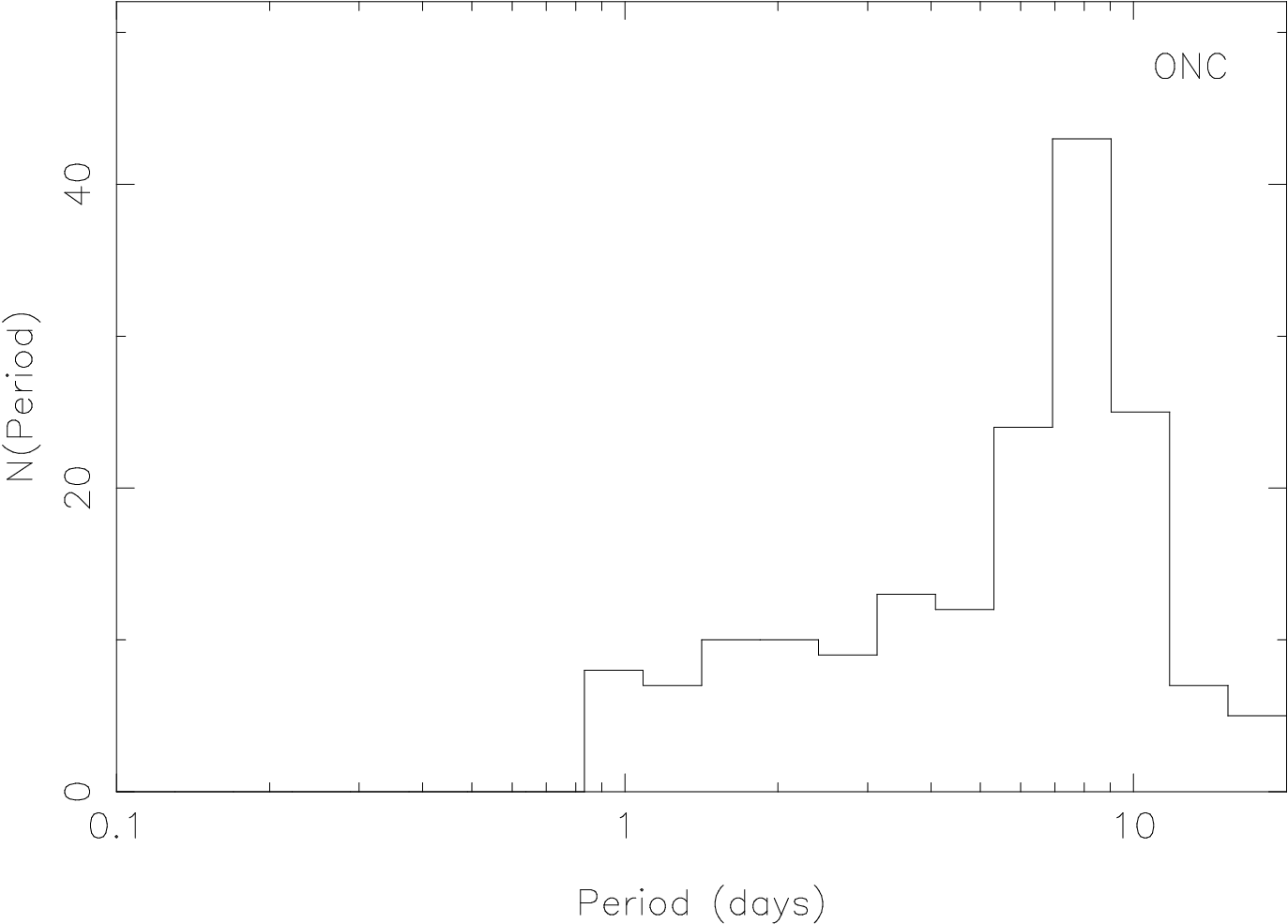}
\caption{Distribution of rotation periods for ONC stars with masses
  $M \ga 0.25\ \msun$ (using the \cite[D'Antona \& Mazzitelli
  1998]{dm98} models; this corresponds to $M \ga 0.4\ \msun$ for the
  \cite[Baraffe et al. 1998]{bcah98} models used in this
  contribution), replotted from \cite{herbst2001}.  The two panels
  show the same distribution binned in linear period (left) and $\log$
  period (right).}
\label{onc_phist}
\end{center}
\end{figure}

In reality, there will be a range of disc lifetimes, and thus the
stars will be ``released'' from this angular velocity regulation at a
range of times, giving rise to a spread of rotation rates on the PMS
and ZAMS, as required by the observations.  Moreover, examining one of
the early-PMS clusters shows us a ``snapshot'' of this process in
action.  The best-observed example is the ONC, shown in Figure
\ref{onc_phist}.

If we presuppose that stars begin with a tight distribution of initial
rotation periods, centred around $8-10\ {\rm days}$, in a histogram of
rotation period we should expect to see a population of stars at this
period, and then a tail of faster-rotating stars, corresponding to the
ones now uncoupled from their discs, with faster rotating stars having
uncoupled earlier.  This is indeed what is seen in Figure
\ref{onc_phist}, where it is instructive to plot the distribution in
$\log$ period rather than the more conventional linear period.

Moreover, this hypothesis makes an observationally-testable
prediction: the population of slowly-rotating stars should show
measurable indications of the presence of a disc (e.g. mid-IR excess)
or of active accretion (i.e. be classical T-Tauri stars, CTTS),
whereas the rapidly-rotating stars should not have discs, or have
recently-dissipated discs, and not be active accretors
(i.e. weak-lined T-Tauri stars, WTTS).

In reality, extracting this observational signal from the cluster data
has proved to be difficult (see \cite[Rebull et al. 2004]{rebull04}
for a detailed discussion of this issue).  \cite{rebull06} presented
the first statistically significant detection of this effect, using
mid-IR excess measurements from the IRAC instrument aboard the
\spitzer\ space telescope, the current method-of-choice, finding that
slowly-rotating stars are indeed more likely to posses discs than
rapidly-rotating stars.  However, in doing so this study revealed a
wrinkle in the argument: they also found a puzzling population of
slowly-rotating stars without discs.  Several subsequent studies
(e.g. \cite[Cieza \& Baliber 2007]{cieza2007}) have confirmed these
results, which now appear to be placed on a firm footing.

\subsection{Main sequence angular momentum losses}

The mechanism for angular momentum loss on the main sequence, at least
for solar mass stars, is thought to be a solar-type magnetised stellar
wind.  The time-dependence of rotation rates in this age range has
been firmly established observationally for solar-type stars since
\cite{kraft67}, and from the age of the Hyades to the Sun, obeys
well the famous \cite{skumanich72} $t^{1/2}$ law (i.e. $\omega
\propto t^{-1/2}$).  This can be reproduced on a more theoretically
motivated framework from parametrised angular momentum loss laws
(usually based on \cite[Kawaler 1988]{kawaler88}; the $t^{1/2}$ law
can be reproduced by setting $n = 3/2$ and $a = 1$ in his model).

Although the \cite{skumanich72} law provides a good description of the
evolution of solar-type stars from the age of the Hyades to the Sun,
and for the majority of stars in the Pleiades to the age of the
Hyades, simply evolving the Pleiades distribution forward in time to
the age of the Hyades using this law produces stars which are rotating
much more rapidly than observed in the Hyades.  Therefore, most
modellers modify the \cite{kawaler88} formalism to incorporate
saturation of the angular momentum losses above a critical angular
velocity $\omega_{\rm sat}$ (\cite[Stauffer \& Hartmann 1987]{plev2};
\cite[Barnes \& Sofia 1996]{bs96}).  The saturation is further assumed
to be mass-dependent, to account for the mass-dependent spin-down
timescales observed on the early-MS.

The resulting angular momentum loss law assumed in our models is:

\begin{equation}
\left({dJ\over{dt}}\right)_{\rm wind} = \left\{ \begin{array}{r}
-K\ \omega^3\ \left({R \over{R_\odot}}\right)^{1/2} \left({M \over{M_\odot}}\right)^{-1/2}, \omega < \omega_{\rm sat} \\
-K\ \omega\ \omega_{\rm sat}^2\ \left({R \over{R_\odot}}\right)^{1/2} \left({M \over{M_\odot}}\right)^{-1/2}, \omega \ge \omega_{\rm sat}
\end{array} \right.
\label{djdt_eqn}
\end{equation}

We note that although saturation is reasonably well-motivated
observationally (e.g. by the saturation observed in the relationship
between X-ray activity and rotation rate), this is a rather
unsatisfactory feature of the present models, since there is a certain
amount of arbitrariness in the way this parameter is introduced
(for example, the choice of power of $\omega$ in Eq. \ref{djdt_eqn}
for the saturated angular velocity dependence of $dJ/dt$ is
arbitrary).  An important area for future work is to develop a
theoretical framework for this phenomenon, and for solar-type winds in
general.

\subsection{Comparison of models to data}
\label{models_section}

The models of rotational evolution we use are described in detail
in \cite{bfa97}, \cite{a98}, \cite{i2007} and \cite{i2007t}.  We
summarise only the most salient features in this contribution.

The methodology of \cite{bfa97} which we adopt here essentially
implements a rotational evolution model by using a standard
non-rotating PMS stellar evolution track to compute the variation of
the stellar parameters as a function of time,  principally the moments
of inertia of the radiative and convective regions of the star, which
are then used as an input to determine the angular velocity as a
function of time.

This model in the simple form described has four parameters (which
could all be functions of mass): the initial angular velocity
$\omega_0$, normalisation of the solar-type angular momentum loss law
$K$ (see Eq. \ref{djdt_eqn}), saturation velocity $\omega_{\rm sat}$,
and the lifetime of the circumstellar disc $\tau_{\rm disc}$.  For
solar-type stars, we can fix $K$ by requiring that the model
reproduces the observed rotation rate of the Sun.

\begin{figure}[h]
\begin{center}
\includegraphics[width=3.2in,angle=270]{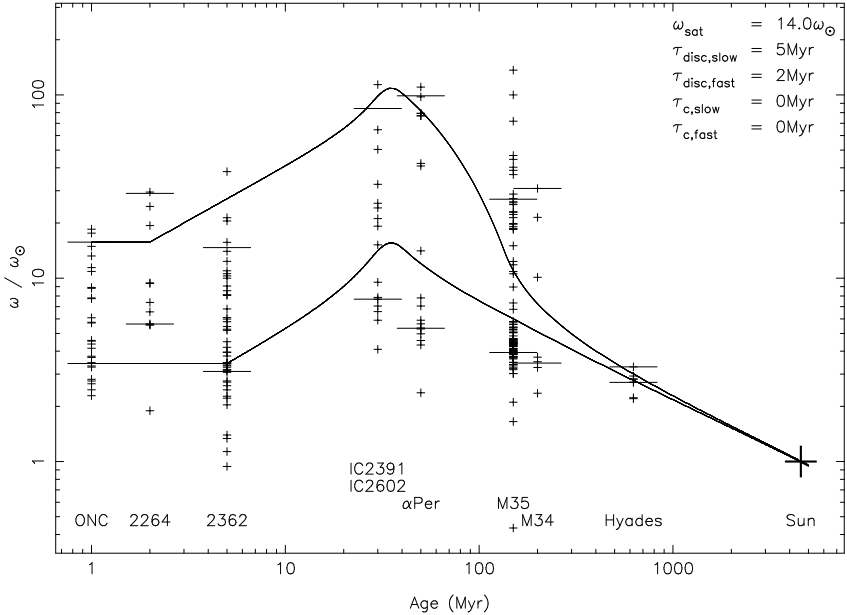}
\caption{Rotational angular velocity $\omega$ plotted as a function of
  time for stars with masses $0.9 < M/\msun \le 1.1$.  Crosses show
  the open cluster rotation period data such that each cluster
  collapses into a vertical stripe on the diagram, and short 
  horizontal lines show the $25$th and $90$th percentiles of $\omega$,
  used to characterise the slow and fast rotators respectively.
  The lines show rotational evolution models for $1.0\ \msun$ stars,
  fit to the percentiles for each cluster using a simple unweighted
  least squares method.  For this plot, we have assumed the stars
  rotate as solid bodies (i.e. constant $\omega$ as a function of
  radius inside the star).  Plotted are the ONC, NGC 2264, NGC 2362,
  IC 2391, IC 2602, $\alpha$ Per, M35, M34, the Hyades, and the Sun.} 
\label{models_1_0_sb}
\end{center}
\end{figure}

Figure \ref{models_1_0_sb} shows the result of fitting this model to
the rotation period data using a simple nonlinear least squares
routine.  For the moment, we have required the radiative core and
convective envelope to have the same angular velocity, i.e. a ``solid
body'' model.

The results show that the model does a reasonable job for the
late-time evolution from the Hyades to the Sun, but this is largely by
construction since we used the \cite{skumanich72} law as an input!  At
earlier times, in general, we find that the rapid rotators are
better-reproduced by the model, with the only major issue being at
$\sim 150\ {\rm Myr}$ where the model underpredicts the upper bound to
the observed rotation rates.  For the slow rotators, the model rotates
too rapidly given the ONC and NGC 2362 as an initial condition around
the ZAMS and early-MS.  The disc lifetimes behave as expected, being
shorter for the rapid rotators.

The difficulty fitting the slow rotators motivates the introduction
of an additional parameter into the model.  In particular, we now
relax the assumption of solid body rotation, and allow the core and
envelope to have different angular velocities, coupling angular
momentum between them on a timescale $\tau_c$ (this is done in a
fashion which tries to equalise their angular velocities).  $\tau_c =
0$ represents the solid body case already considered.  The net effect
of this ``core-envelope decoupling'' from the point of view of the
observed rotation rate (which is that of the envelope) is to ``hide''
angular momentum in the core when it forms, which is then coupled back
into the envelope gradually, providing a ``late-time replenishment''
of the surface angular velocity.

\begin{figure}[h]
\begin{center}
\includegraphics[width=3.2in,angle=270]{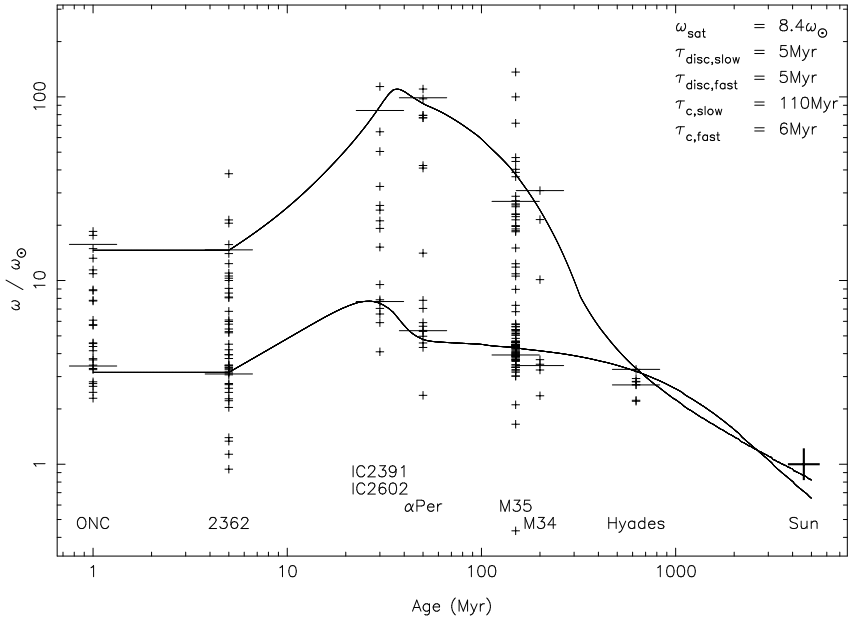}
\caption{As Figure \ref{models_1_0_sb}, but now allowing the radiative
  core and convective envelope of the stars to have different angular
  velocities, coupling angular momentum between the two on a timescale
  $\tau_c$.  We have also omitted NGC 2264, since the models presently
  have difficulty explaining the observations in this cluster.}
\label{models_1_0}
\end{center}
\end{figure}

Figure \ref{models_1_0} shows the result of applying this revised
model to the solar-type stars.  We can see that the fit is
substantially improved.  Importantly, we find that the value of
$\tau_c$ for the slow rotators is relatively large, $\sim 110\ {\rm
  Myr}$ indicating inefficient core-envelope coupling, whereas for the
fast rotators it is short, $\sim 6\ {\rm Myr}$: these stars have
efficient core-envelope coupling and rotate more like solid bodies.  A
corollary of this is that slow rotators develop a higher degree of
rotational shear across the convective/radiative boundary.

This result may have important observationally-testable consequences.
\cite{bouvier2008} discusses one of these: the impact on the surface
abundance of lithium.  In particular, the higher rotational shear in
the slow rotators is likely to induce additional mixing in these
stars, which may bring Li down into the star where it can be burnt.
The net effect would be a higher level of lithium depletion in slow
rotators than fast rotators, and these should therefore display a
lower lithium abundance on the main sequence.  Indeed, \cite{plev3}
report that this appears to be the case in the Pleiades.

\begin{figure}[h]
\begin{center}
\includegraphics[width=3.2in,angle=270]{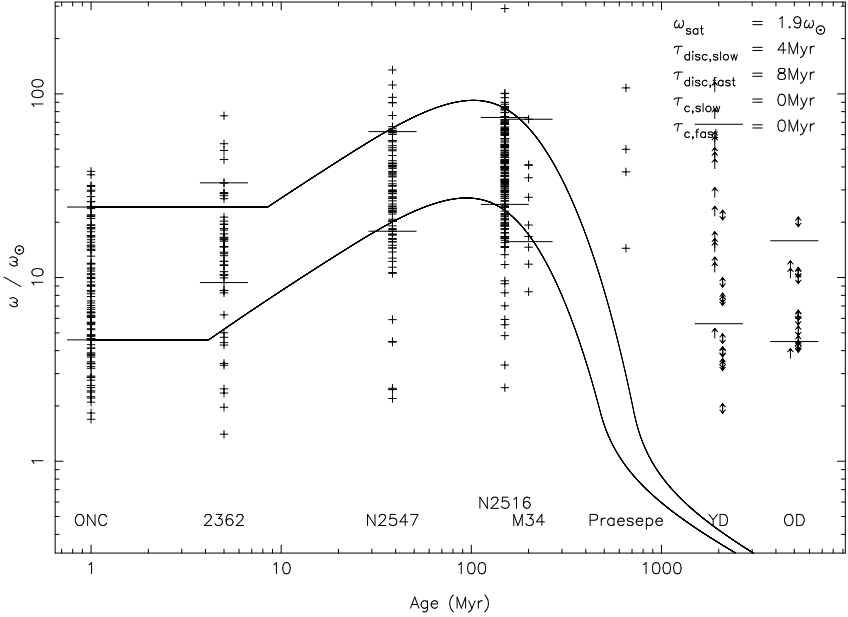}
\caption{As Figure \ref{models_1_0}, except for stars with masses $0.1
  < M/\msun \le 0.4$, and models for $0.25\ \msun$.  Data for NGC
  2547, NGC 2516 and Praesepe are shown in addition to those clusters
  already included in Figure \ref{models_1_0}.  We have also included
  lower limits from $v \sin i$ measurements for field M-dwarfs from
  \cite{delfosse98}, placing the young disc population at $2\ {\rm
  Gyr}$ and the old disc at $5\ {\rm Gyr}$ (e.g. \cite[Feltzing \&
  Bensby 2008]{feltzing2008}; although note that in reality, these
  populations have a wide dispersion in age).  Many of their points are only
  upper limits in $v \sin i$, and are shown as double-headed arrows in
  the figure, plotted offset slightly from the other data in
  age for clarity.  Note the sparse coverage of
  cluster data, and particularly rotation periods, at the oldest ages.
  This is largely due to the intrinsic faintness of old, mid$-$late
  M-dwarfs.  Several proposed surveys aim to fill in this region of
  parameter space within the next couple of years.}
\label{models_0_25}
\end{center}
\end{figure}

Having examined the models for solar-type stars, we now proceed to the
fully-convective, very low-mass stars ($M \la 0.4\ \msun$).  Figure
\ref{models_0_25} shows the results of applying the same models to
this mass domain.  Since these stars are fully-convective, there is no
possibility for core-envelope decoupling, so they should rotate
as solid bodies.

The model is clearly a poor match to the observations.  In particular,
the solar-type wind appears to loose too much angular momentum at
late-times, resulting in a sharp drop in the predicted rotation rates
starting at $\sim 200\ {\rm Myr}$, which does not appear to be seen,
although the available data are still very limited at present, with $v
\sin i$ measurements giving only upper limits for the slowest rotating
stars.  This cannot be ``fixed'' by invoking saturation, since this
has already been done for the model shown in Figure \ref{models_0_25},
which has $\omega_{\rm sat} = 1.8\ \omega_\odot$ (i.e. every star is
saturated in the relevant age range).

It would not be surprising that a solar-type wind does not reproduce
the evolution of fully-convective stars, since this is thought to be
launched at a tachocline of shear at the radiative/convective
interface, and these stars posses none.  However, as highlighted by
J. Stauffer in the questions, we should be cautious in
overinterpreting the rather sparse data presently available at these
masses and ages, especially given the large number of $v \sin i$ upper
limits and the natural bias of the rotation period method toward more
active, and hence presumably more rapidly-rotating, stars, resulting
from the extremely small photometric amplitudes of the rotational
modulations at these old ages.

This domain of very low-masses and late ($\ga 500\ {\rm Myr}$) ages
represents an important area for future observational studies to
explore, and could provide us with a valuable insight into the winds
of fully-convective stars.

\section*{Acknowledgments}

This research has made use of the SIMBAD database, operated
at CDS, Strasbourg, France, the WEBDA database, operated at the
Institute for Astronomy of the University of Vienna, and the Open Cluster 
Database, as provided by C.F. Prosser and J.R. Stauffer, which may
currently be accessed at {\tt
  http://www.noao.edu/noao/staff/cprosser/}, or by anonymous ftp to
{\tt 140.252.1.11}, {\tt cd /pub/prosser/clusters/}.  We thank
Isabelle Baraffe for providing the stellar evolution tracks used to
compute the models in \S \ref{models_section}.


\begin{thebibliography}{}

\bibitem[Allain et al. (1996)]{aper7}
{Allain, S., Fernandez, M., Mart\'{i}n, E. L., \& Bouvier, J.} 1996,
\textit{A\&A}, 314, 173

\bibitem[Allain (1998)]{a98}
{Allain, S.} 1998,
\textit{A\&A}, 333, 629

\bibitem[Baraffe et al. (1998)]{bcah98}
{Baraffe, I., Chabrier, G., Allard, F., \& Hauschildt, P. H.} 1998,
\textit{A\&A}, 337, 403

\bibitem[Barnes et al. (1999)]{bsps99}
{Barnes, S. A., Sofia, S., Prosser, C. F., \& Stauffer, J. R.} 1999,
\textit{ApJ}, 516, 263

\bibitem[Barnes \& Sofia (1996)]{bs96}
{Barnes, S. \& Sofia, S.} 1996,
\textit{ApJ}, 462, 746

\bibitem[Barnes et al. (1998)]{aper11}
{Barnes, J. R., Collier Cameron, A., Unruh, Y. C., Donati, J. F.,
  \& Hussain, G. A. J.} 1998,
\textit{MNRAS}, 299, 904

\bibitem[Bouvier et al. (1997)]{bfa97}
{Bouvier, J., Forestini, M., \& Allain, S.} 1997,
\textit{A\&A}, 326, 1023

\bibitem[Bouvier (2008)]{bouvier2008}
{Bouvier, J.} 2008,
\textit{A\&A}, 489, 53

\bibitem[Cieza \& Baliber (2006)]{cieza2006}
{Cieza, L. \& Baliber, N.} 2006,
\textit{ApJ}, 649, 862

\bibitem[Cieza \& Baliber (2007)]{cieza2007}
{Cieza, L. \& Baliber, N.} 2007,
\textit{ApJ}, 671, 605

\bibitem[Cohen et al. (2004)]{cohen2004}
{Cohen, R. E., Herbst, W., \& Williams, E. C.} 2004,
\textit{AJ}, 127, 1602

\bibitem[Collier Cameron et al. (1995)]{cc95}
{Collier Cameron, A., Campbell, C. G., \& Quaintrell, H.} 1995,
\textit{A\&A}, 298, 133

\bibitem[D'Antona \& Mazzitelli (1998)]{dm98}
{D'Antona, F., \& Mazzitelli, I.} 1998,
in R. Rebolo, E. L. Mart\'{\i}n, \& M. R. Zapatero Osorio (eds.),
\textit{Brown dwarfs and extrasolar planets}, ASP Conf. Series 134, p.\, 442.

\bibitem[Delfosse et al. (1998)]{delfosse98}
{Delfosse, X., Forveille, T., Perrier, C., \& Mayor, M.} 1998,
\textit{A\&A}, 331, 581

\bibitem[Feltzing \& Bensby (2008)]{feltzing2008}
{Feltzing, S., \& Bensby, T.} 2008,
in P. Barklem, A. Korn, \& B. Plez (eds.),
\textit{A stellar journey}, Physica Scripta, in press ({\tt arXiv:0811.1777})

\bibitem[Haisch et al. (2001)]{haisch2001}
{Haisch, K. E., Lada, E. A., \& Lada, C. J.} 2001,
\textit{ApJ}, 553, 153

\bibitem[Hartman et al. (2008)]{hartman2008}
{Hartman, J. D., et al.} 2008,
\textit{ApJ}, in press ({\tt arXiv:0803.1488})

\bibitem[Herbst et al. (2001)]{herbst2001}
{Herbst, W., Bailer-Jones, C. A. L. \& Mundt, R.} 2001,
\textit{ApJ}, 554, 197

\bibitem[Herbst et al. (2002)]{herbst2002}
{Herbst, W., Bailer-Jones, C. A. L., Mundt, R., Meisenheimer, K., \&
  Wackermann, R.} 2002,
\textit{A\&A}, 396, 513

\bibitem[Irwin et al. (2006)]{i2006}
{Irwin, J., Aigrain, S., Hodgkin, S., Irwin, M., Bouvier, J., Clarke,
  C., Hebb, L., \& Moraux, E.} 2006,
\textit{MNRAS}, 370, 954

\bibitem[Irwin et al. (2007)]{i2007}
{Irwin, J., Hodgkin, S., Aigrain, S., Hebb, L., Bouvier, J., Clarke,
  C., Moraux, E. \& Bramich D. M.} 2007,
\textit{MNRAS}, 377, 741

\bibitem[Irwin (2007)]{i2007t}
{Irwin, J.} 2007,
Ph.D. thesis, University of Cambridge

\bibitem[Irwin et al. (2008a)]{i2008a}
{Irwin, J., Hodgkin, S., Aigrain, S., Bouvier, J., Hebb, L., \& Moraux, E.} 2008a,
\textit{MNRAS}, 383, 1588

\bibitem[Irwin et al. (2008b)]{i2008b}
{Irwin, J., Hodgkin, S., Aigrain, S., Bouvier, J., Hebb, L., Irwin,
  M., \& Moraux, E.} 2008b,
\textit{MNRAS}, 384, 675

\bibitem[Irwin et al. (2009)]{i2009}
{Irwin, J., Aigrain, S., Bouvier, J., Hebb, L., Hodgkin, S., Irwin,
  M., \& Moraux, E.} 2009,
\textit{MNRAS}, in press ({\tt arXiv:0810.5110})

\bibitem[Jeffries et al. (1998)]{jeffries98}
{Jeffries, R. D., James, D. J., \& Thurston, M. R.} 1998,
\textit{MNRAS}, 300, 550

\bibitem[Jones et al. (1996)]{plev4}
{Jones, B. F., Fischer, D. A., \& Stauffer, J. R.} 1996,
\textit{AJ}, 112, 1562

\bibitem[K\"{o}nigl (1991)]{konigl91}
{K\"{o}nigl, A.} 1991,
\textit{ApJ}, 370, L37

\bibitem[Kawaler (1988)]{kawaler88}
{Kawaler, S. D.} 1998,
\textit{ApJ}, 333, 236

\bibitem[Kraft (1967)]{kraft67}
{Kraft, R. P.} 1967,
\textit{ApJ}, 150, 551

\bibitem[Krishnamurthi et al. (1998)]{ple7}
{Krishnamurthi, A., et al.} 1998,
\textit{ApJ}, 493, 914

\bibitem[Lamm et al. (2005)]{lamm2005}
{Lamm, M. H., Mundt, R., Bailer-Jones, C. A. L., \& Herbst, W.} 2005,
\textit{A\&A}, 430, 1005

\bibitem[Littlefair et al. (2005)]{little2005}
{Littlefair, S. P., Naylor, T., Burningham, B., \& Jeffries, R. D.} 2005,
\textit{MNRAS}, 358, 341

\bibitem[Magnitskii (1987)]{ple3}
{Magnitskii, A. K.} 1987,
\textit{Soviet Astron.} (Letters), 13, 451

\bibitem[Makidon et al. (2004)]{makidon2004}
{Makidon, R. B., Rebull, L. M., Strom, S. E., Adams, M. T., \& Patten,
  B. M.} 2004,
\textit{AJ}, 127, 2228

\bibitem[Mart\'{\i}n \& Zapatero Osorio (1997)]{aper8}
{Mart\'{\i}n, E. L. \& Zapatero Osorio, M. R.} 1997,
\textit{MNRAS}, 286, L17

\bibitem[Matt \& Pudritz (2005)]{mp2005}
{Matt, S. \& Pudritz, R. E.} 2005,
\textit{ApJ}, 632, 135

\bibitem[Meibom et al. (2008)]{meibom2008}
{Meibom, S., Mathieu, R. D., \& Stassun K. G.} 2008,
\textit{ApJ}, in press ({\tt arXiv:0805.1040})

\bibitem[O'Dell \& Collier Cameron (1993)]{aper4}
{O'Dell, M. A. \& Collier Cameron, A.} 1993,
\textit{MNRAS}, 262, 521

\bibitem[O'Dell et al. (1994)]{aper5}
{O'Dell, M. A., Hendry, M. A., \& Collier Cameron, A.} 1994,
\textit{MNRAS}, 268, 181

\bibitem[O'Dell at al. (1996)]{aper6}
{O'Dell, M. A., Hilditch, R. W., Collier Cameron, A. \& Bell, S. A.} 1996,
\textit{MNRAS}, 284, 874

\bibitem[Patten \& Simon (1996)]{ps96}
{Patten, B. M. \& Simon, T.} 1996,
\textit{ApJS}, 106, 489

\bibitem[Prosser (1991)]{aper3}
{Prosser, C. F.} 1991,
Ph.D. Thesis, University of California, Santa Cruz

\bibitem[Prosser et al. (1993a)]{ple4}
{Prosser, C. F., Schild, R. E., Stauffer, J. R., Jones, B. F.} 1993a,
\textit{PASP}, 105, 269

\bibitem[Prosser et al. (1993b)]{ple5}
{Prosser, C. F., et al.} 1993b,
\textit{PASP}, 105, 1407

\bibitem[Prosser et al. (1995)]{ple6}
{Prosser, C. F., et al.} 1995,
\textit{PASP}, 107, 211

\bibitem[Prosser \& Randich (1998)]{aper9}
{Prosser, C. F. \& Randich, S.} 1998,
\textit{AN}, 319, 210

\bibitem[Prosser et al. (1998)]{aper10}
{Prosser, C. F., Randich, S. \& Simon, T.} 1998,
\textit{AN}, 319, 215

\bibitem[Queloz et al. (1998)]{plev5}
{Queloz, D., Allain, S., Mermilliod, J.-C., Bouvier, J. \& Mayor, M.} 1998,
\textit{A\&A}, 335, 183

\bibitem[Radick et al. (1987)]{hya1}
{Radick, R. R., Thompson, D. T., Lockwood, G. W., Duncan, D. K., \&
  Baggett, W. E.} 1987,
\textit{ApJ}, 321, 459

\bibitem[Rebull et al. (2004)]{rebull04}
{Rebull, L. M., Wolff S. C., \& Strom S. E.} 2004,
\textit{AJ}, 127, 1029

\bibitem[Rebull et al. (2006)]{rebull06}
{Rebull, L. M., Stauffer, J. R., Megeath, S. T., Hora, J. L., \&
  Hartmann, L.} 2006,
\textit{ApJ}, 646, 297

\bibitem[Scholz \& Eisl\"offel (2004a)]{se2004a}
{Scholz, A. \& Eisl\"offel, J.} 2004,
\textit{A\&A}, 419, 249

\bibitem[Scholz \& Eisl\"offel (2004b)]{se2004b}
{Scholz, A. \& Eisl\"offel, J.} 2004,
\textit{A\&A}, 421, 259

\bibitem[Scholz \& Eisl\"offel (2005)]{se2005}
{Scholz, A. \& Eisl\"offel, J.} 2005,
\textit{A\&A}, 429, 1007

\bibitem[Scholz \& Eisl\"offel (2007)]{se2007}
{Scholz, A. \& Eisl\"offel, J.} 2007,
\textit{MNRAS}, 381, 1638

\bibitem[Skumanich (1972)]{skumanich72}
{Skumanich, A.} 1972,
\textit{ApJ}, 171, 565

\bibitem[Soderblom et al. (1993)]{plev3}
{Soderblom, D. R., Stauffer, J. R., Hudon, J. D., \& Jones, B. F.} 1993,
\textit{ApJS}, 85, 315

\bibitem[Stassun et al. (1999)]{stassun99}
{Stassun, K. G., Mathieu, R. D., Mazeh, T., \& Vrba, F. J.} 1999,
\textit{AJ}, 117, 2941

\bibitem[Stauffer et al. (1984)]{plev1}
{Stauffer, J. R., Hartmann, L., Soderblom, D. R., \& Burnham, N.} 1984,
\textit{ApJ}, 280, 202

\bibitem[Stauffer et al. (1985)]{aper1}
{Stauffer, J. R., Hartmann, L. W., Burnham, J. N., \& Jones, B. F.} 1985,
\textit{ApJ}, 289, 247

\bibitem[Stauffer \& Hartmann (1987)]{plev2}
{Stauffer, J. R. \& Hartmann, L. W.} 1987,
\textit{ApJ}, 318, 337

\bibitem[Stauffer et al. (1987)]{ple2}
{Stauffer, J. R., Schild, R. A., Baliunas, S. L., \& Africano, J. L.} 1987,
\textit{PASP}, 99, 471

\bibitem[Stauffer et al. (1989)]{aper2}
{Stauffer, J. R., Hartmann, L. W., \& Jones, B. F.} 1989,
\textit{ApJ}, 346, 160

\bibitem[Terndrup et al. (1999)]{t99}
{Terndrup, D. M., Krishnamurthi, A., Pinsonneault, M. H., \& Stauffer,
  J. R.} 1999,
\textit{AJ}, 118, 1814

\bibitem[Terndrup et al. (2000)]{plev6}
{Terndrup, D. M., Stauffer, J. R., Pinsonneault, M. H., Sills, A.,
  Yuan, Y., Jones, B. F., Fischer, D., \& Krishamurthi, A.} 2000,
\textit{AJ}, 119, 1303

\bibitem[Terndrup et al. (2002)]{ter2002}
{Terndrup, D. M., Pinsonneault, M., Jeffries, R. D., Ford, A., \&
  Sills, A.} 2002,
\textit{ApJ}, 576, 950

\bibitem[van Leeuwen et al. (1987)]{ple1}
{Van Leeuwen, F., Alphenaar, P, \& Meys, J. J. M.} 1987,
\textit{A\&AS}, 67, 483

\end{thebibliography}
\end{document}